\documentclass{statsoc}
\usepackage{amsmath,amsfonts,graphicx}
\newcommand{\Fig}[1]{Fig.~\ref{#1}}
\newcommand{\Sec}[1]{Section~\ref{#1}}
\newcommand{\Eqn}[1]{Equation~(\ref{#1})}

\def\card{{\rm card\,}}
\def\bra{\langle\,}
\def\ket{\,\rangle}

\title[Ancestral trees for binary trait data]{Dated ancestral trees from binary trait data and its application to
the diversification of languages}

\author[GK Nicholls {\it et al.}]{Geoff K. Nicholls}
\address{Department of Statistics,
         Oxford,
         UK.}
\coaddress{GK Nicholls, Department of Statistics, 1 South Parks Road, Oxford OX1 3TG, UK}
      \email{nicholls@stats.ox.ac.uk}
    \author{Russell D. Gray}
      \address{Department of Psychology, Auckland University,
       Auckland,
       New Zealand.}

\begin{document}

\begin{abstract}
Binary trait data record the presence or absence of distinguishing
traits in individuals. We treat the problem of estimating
ancestral trees with time depth from binary trait data.
Simple analysis of such data is problematic. Each homology class of traits
has a unique birth event on the tree, and the birth event of a trait
visible at the leaves is biased towards the leaves. We propose
a model-based analysis of such data, and present an MCMC algorithm that can sample
from the resulting posterior distribution. Our model is based on using a
birth-death process for the evolution of the elements of sets of traits.
Our analysis correctly accounts for the removal of singleton traits,
which are commonly discarded in real data sets.
We illustrate Bayesian inference for two binary-trait
data sets which arise in historical linguistics.
The Bayesian approach allows for the
incorporation of information from ancestral languages.
The marginal prior distribution of the root time is uniform.
We present a thorough analysis of the robustness of our results to model
mispecification, through analysis of predictive
distributions for external data, and fitting data simulated
under alternative observation models. The reconstructed ages
of tree nodes are relatively robust, whilst posterior probabilities
for topology are not reliable.
\end{abstract}

\section{Introduction}

A great deal of progress has been made on the statistical analysis of
DNA sequence data, and in particular for model-based estimation of
genealogy. No equivalent statistical framework exists for
trait-based cladistics. However, qualitative and
quantitative trait data may be used to recover
dated tree-like histories in situations where we have no genetic sequence data.
Progress is possible when the traits are similar in type,
so that some unifying assumption about their evolution is justified.


We give statistical methodology for tree-estimation from
binary trait data. These data are made up of
binary sequences, each sequence recording for one taxon the presence or absence of a list
of traits. Pigeon wings and sparrow wings are {\it instances} of the trait ``bird wings''
displayed at the taxa ``Pigeon" and ``Sparrow". In our model, two instances of a trait
are necessarily {\it homologous}, that is, they descend from a common ancestor.
Trait observation models have many missing data. Birth
times of observed traits are unknown, and traits displayed at less than two taxa
may be discarded. We model these missing data, integrate them
out of the analysis analytically, and measure the random error
using sample based Bayesian inference.

The outline of this paper is as follows. In the first half
(Sections~\ref{sec:model}~to~\ref{sec:mcmc}) we set up Bayesian inference
for a class of trait models. We give observation models,
likelihood evaluation, prior, posterior and a MCMC scheme. In the second half of the
paper (Sections~\ref{sec:data}~to~\ref{sec:mm}) we apply the inference scheme
to two closely related data sets. We review previous studies of these data
and describe the particular models we fit, then present
results and model mispecification analysis. Readers interested in the application only
should read the first paragraph of \ref{sec:model}, and all of Section \ref{sec:treeprior},
before jumping to the data analysis in Sections~\ref{sec:data}~and~\ref{sec:inference}.
Graphics illustrating this application make up the bulk of the supplement \cite{nicholls07},
see \verb?http://www.stats.ox.ac.uk/~nicholls/linkfiles/papers/NichollsGray06-SUPP.pdf?.
Supplement section labels correspond to the section labels in this paper.

We begin in \Sec{sec:model} with the observation process.
In \Sec{sec:lkd} we give an efficient scheme for evaluating the corresponding likelihood.
The model, described in \cite{huson04} and \cite{atkinson05},
is the natural stochastic process representing Dollo's parsimony criterion,
since each instance of a given trait descends from a single innovation.
In \cite{huson04} the traits are distinct genes which are present or absent in an individual,
and trees are built using a maximum-likelihood pairwise-distance,
and the neighbor-joining methods of \cite{saitou87}. 
Our own work has been motivated by a pair of data-sets,
\cite{dyen97} and \cite{ringe02}, recording trait presence and absence for Indo-European languages.
Here traits are {\it cognate classes} (also called {\it lexical traits}), that is, homology classes of words of closely similar meaning.
Thus English, Flemish and Danish share the trait ``all/alle/al'' whilst
Spanish, Catalan and Italian lack that trait, but share ``todo/tot/tutto''.

In \Sec{sec:treeprior} we write down two prior probability distributions for trees.
The first imposes an exponential penalty on branch length. The second is designed to
be non-informative with respect to the height of the root node, and otherwise uniform
on the space of trees. This distribution is a new class of tree priors and is likely to be useful
in a broader phylogenetic setting. General calibration constraints are introduced in \Sec{sec:treeprior}
with specific examples in \Sec{sec:caldata}.
These constraints, derived from historical records, bound some tree node ages above and below
and are used to fix trait birth and death rates. They determine the parameter space of trees.
In \Sec{sec:cal} we gather together the results of \Sec{sec:lkd} and \Sec{sec:treeprior}
and write down the full posterior density for the model parameters. We give closed form results
for the two leaf case, and verify that we reproduce the distance measure of \cite{huson04}.
We give a brief description of our MCMC algorithm for sampling the posterior density
in \Sec{sec:mcmc}.

In \Sec{sec:data} we introduce the cognate data and the associated
calibration constraint data and summarize previous work. \Sec{sec:inference}
has two parts. \Sec{sec:fitted} gives further details of the model we fit
and \Sec{sec:results} summarizes results and conclusions.

Statistical contributions to
the dating of language branching events have been rejected by linguists.
Dating efforts are criticized for their assumption of a constant
rate of language change at all times and in all places, the so called ``glottal clock''.
\cite{bergsland62} found examples of extreme rates, but employed counter-examples biased
by data-selection. \cite{blust00} links rate heterogeneity to long-branch attraction.
However, neither criticsm considers even the random component of the error. In this respect
we are repeating the comments of \cite{sankoff73}.
In \Sec{sec:mm} we investigate model errors. Although we find evidence for model mispecification
we nevertheless reproduce, to within random error, age estimates in
analyses across near-independent data, and in reconstructions from synthetic data simulated
under likely model-violation scenarios.

 \section{A model of binary trait evolution}
 \label{sec:model}
In this section we specify an observation model for traits evolving on a fixed tree.
We preface this section with a qualitative description of the model.
An individual is represented by a set of traits which are distinguishable and non-interacting.
As calendar time passes, new traits are born into the set at a constant rate.
Each such birth generates a new homology class of trait instances.
At a tree node, two identical copies of the set
entering the node leave that node. Two instances of each trait entering a node leave the node, one
instance in each set. Each instance of each trait in each set dies independently
at a constant rate. A graphical illustration of the process and notation is
given in the supplement, \cite{nicholls07}.
Models of the tree itself are given in Section~\ref{sec:treeprior}.

We begin our formal description with notation for the tree. Let $g=(E,V,t)$ be a rooted
binary tree with $L$ leaves plus one extra node (label $R^*$) ancestral to the root itself. The set of all nodes is
then $V=\{1,2,\ldots,2L\}$, with leaf nodes $V_L$, ancestral nodes $V_A$ and edges
$\bra i,j\ket\in E$ where $i,j\in V$ and $i<j$. Node ages
$t=(t_1,t_2,\ldots,t_{2L})$ are ordered $t_i\le t_{i+1}$ and increase from the leaves to the root.
The root node label is $R=2L-1$ and there is an additional node $R^*=2L$ with age
$t_{R^*}=\infty$ which is connected to the root via an edge $\bra R,R^*\ket\in E$. Our convention is $R^*\not\in V_A$,
so $V=\{R^*\}\cup V_A\cup V_L$. Leaves may be staggered in time.

Next we describe the evolution of traits. 
Sets of trait instances are sets of trait labels
which evolve along the branches of $g$ from the root towards the leaves,
in the direction of decreasing age.
Identify edge $\bra i,j\ket$ by the node $i$ at the base of that edge
(edges are directed for increasing age, from leaf to root,
in the opposite direction to the
calendar evolution of traits, from root to leaf).
For $\bra i,j\ket\in E$ and $\tau\in[t_i,t_j)$ denote by $(\tau,i)$
a time point on a branch of $g$ and by
\[ [g]=\bigcup_{\bra i,j\ket\in E}\bigcup_{\tau\in [t_i,t_j)} \{(\tau,i)\},\]
the set of all such points, including
points on the edge $\bra R,R^*\ket$ of infinite length.
For each branch $\bra i,j\ket\in E$ define a set-valued process $H(\tau,i)=\{h_1,h_2,\ldots,h_{N(\tau,i)}\}$
of trait labels $h_a\in \mathbb{Z},\ a=1,2,\ldots,N(\tau,i)$. The elements of
$H(\tau,i)$ are realized by a simple reversible birth-death process which acts along
each edge of the tree. Set elements are born
at constant rate $\lambda$. The label for each new born trait is
unique to that trait, but otherwise arbitrary.
Set elements die at constant {\it per capita} rate $\mu$.
At a branching event $(t_i,i)\in[g]$,
set $H(t_i,i)$ is copied onto the top of the two branches $\bra j,i\ket$ and $\bra k,i\ket$
emerging from $i$, so that the evolution of $H(\tau,j)$ and $H(\tau,k)$ is conditional
on $H(t_i,j)=H(t_i,k)=H(t_i,i)$.

The number of elements $N(t_R,R)$ in the root set is the number of traits
born in $[t_R,\infty)$ which survive to time $t_R$. These surviving traits
are generated at rate $\lambda(\tau,R)=\lambda\exp(-\mu (\tau-t_R))$, so their number
is Poisson, mean $\lambda/\mu$. 
The process may therefore be initialized by simulating $N(t_R,R)\sim\Pi(\lambda/\mu)$,
and assigning $N(t_R,R)$ arbitrary trait labels $H(t_R,R)=\{1,2,\ldots,N(t_R,R)\}$
to the root set.

The data $D^{(L)}=(H_i, i\in V_L)$, where
\begin{equation}
\label{eq:noabsent}
H_i=H(t_i,i), \quad i\in V_L
\end{equation}
are an ordered list of the sets of trait labels observed at the tree leaves.
Suppose that in all there are $N$
distinct trait labels $C=(c_1,c_2,\ldots,c_N)$ (so, $C=\left(\cup_{i\in V_L} H_i\right)$)
displayed at the leaves.
We can represent the data as $N$ sets of taxa labels also,
with set $M_a$ giving the leaves at which trait $c_a$ appears. This representation is
$D^{(N)}=(M_1, M_2,\ldots,M_N)$, with $M_a=\{i: c_a\in H_i,\ i\in V_L\}$ for $a=1,\ldots,N$.

Traits displayed at just one taxon are often dropped from the data.
It is argued that these
singleton traits do not inform tree topology. This is
not the case in the model we have described,
since singleton traits are informative of time depth. Referring to the data analyzed in \Sec{sec:data},
\cite{gray03} drop singleton traits in their binary registration of the \cite{dyen97}
data-set but retain them in their registration of the \cite{ringe02} data.
Let $\mathbb{I}_{c\in H(t_j,j)}=1$ if $c\in H(t_j,j)$ and zero otherwise.
The thinned data is $D^{(L)}=(H_i, i\in V_L)$, with
\begin{equation}
\label{eq:nounique}
H_i=\left\{c\in H(t_i,i): \sum_{j\in V_L} \mathbb{I}_{c\in H(t_j,j)}>1 \right\},\quad i\in V_L.
\end{equation}
We call this observation model, which drops singleton traits, NOUNIQUE,
in contrast to the model NOABSENT defined by \Eqn{eq:noabsent}.
We write $D$ for generic NOABSENT or NOUNIQUE data.


\cite{felsenstein92} gives the likelihood for a
Poisson process acting on a finite state space,
along the branches of a tree, conditioned to show states other than the zero
state at the leaves.
\cite{lewis01} proposes applying certain trait models of this kind (so-called Jukes-Cantor models)
to morphological character data,
in a maximum likelihood analysis. \cite{lewis01} mentions the problem
of thinning traits displayed at a single taxon, and treats it by ensuring the data
are not so thinned. \cite{nylander04} fit models from the same family, allowing
for the thinning of all parsimony
uninformative characters (traits displayed at $0$, $1$, $L-1$ or $L$ leaves).
These models do not constrain a trait to be generated at a single birth event.
The authors model a fixed number of traits which move back and forward between different categorical
values indefinitely. The number of distinct traits is fixed for all time.
We impose a single birth event for a trait and an evolution which proceeds
from absence to presence to absence only. The number of distinct traits
generated by our process is random, so that the total number is informative of the relative rates of birth and death.
The model we have described resembles the \cite{watterson75} infinite sites model,
but here trait-death is in effect back-mutation. Our model is similar to
the infinite alleles model of \cite{kimura64}, though the number of alleles is not random,
whilst the number of traits is random.

\section{Likelihood calculations}
\label{sec:lkd}

The likelihood for $g,\mu$ and $\lambda$ is given in terms of the distribution of
the point process of birth points for those traits displayed in the data.
Let $X=\{X_1,X_2,\ldots,X_N\}$ be a random set of trait birth-points in $[g]$.
The Poisson process generating $X$ is obtained by thinning realizations of
a constant rate process. Suppose a trait with label
$c$ is born at $z\in [g]$; let $O(z)=\sum_{i\in V_L} \mathbb{I}_{c\in H_i}$
give the number of taxa displaying trait $c$ (after any thinning).
If $\Pr\{O(z)>d|z,g,\mu\}$ is the probability for a trait, born at $z\in[g]$
to appear in the data at $d+1$ or more leaves,
then the trait birth-rate at $z$ in process $X$ is $\lambda(z)=\lambda\Pr\{O(z)>d|z,g,\mu\}$,
where $d=0$ under the NOABSENT observation model and $d=1$ under NOUNIQUE.

The distribution of $X$ is defined on the space $\mathcal{X}$ of all finite subsets $x\subset [g]$.
For $f:[g]\rightarrow \Re$, define the integral $\int_{[g]} f(z) dz$ along tree branches by
\[
\int_{[g]} f(z)dz= \sum_{\bra i,j\ket\in E} \int_{t_i}^{t_j} f\left((\tau,i)\right)\,d\tau.
\]
Now, suppose $X=x$ with $x=\{x_1,x_2,\ldots,x_N\}$
and $x_a=(\tau_a,i_a)$ for $a=1,\ldots,N$, so that $x_a\in [g]$ identifies the point on the tree where
trait $c_a$ was born. Let $dx_a=dz$ at $x_a=z$.
The density of the random set $X=x$, with respect to $dx=dx_1dx_2 \ldots dx_N$ on $\mathcal{X}$, is
\[
f_X(x|g,\mu,\lambda)=\exp\left(-\int_{[g]} \lambda(z)dz\right)\prod_{a=1}^N \lambda(x_a).
\]
The total number of distinct traits in the data $N\sim \Pi\left(\int_{[g]} \lambda(z)dz\right)$
has a Poisson distribution.

Trait birth points are nuisance parameters, which we integrate out of the likelihood
under the density $f_X$. 
Denote by $\Pr\{M_a=m_a|x_a,g,\mu,O(x_a)>d\}$
the probability for a trait, born
at $x_a$, to be displayed at the leaves listed in set $m_a$ and no others, conditional on being
displayed in at least $d$ leaves.
The likelihood, $P(D|g,\mu,\lambda)$, is
\begin{eqnarray*}
  P(D|g,\mu,\lambda) &=& \int_{\mathcal X} P(D|x,g,\mu)f_X(x|g,\mu,\lambda)dx \\
    &=& \frac{e^{-\int_{[g]} \lambda(z)dz}}{N!} \prod_{a=1}^N \lambda \int_{[g]}\Pr\{M_a=m_a|x_a,g,\mu,O(x_a)>d\}\Pr\{O(x_a)>d|x_a,g,\mu\}dx_a.
\end{eqnarray*}
The outcome $\{M_a=m_a,O(x_a)>d\}$ is identical to the
outcome $\{M_a=m_a\}$ for traits in the data, since those traits
already satisfy the thinning condition $\card m_a>d$ for each $a=1,2,\ldots,N$. It follows that events in the data satisfy
\[
\Pr\{M_a=m_a|x_a,g,\mu,O(x_a)>d\} = \frac{\Pr\{M_a=m_a|x_a,g,\mu\}}{\Pr\{O(x_a)>d|x_a,g,\mu\}},
\]
and consequently the likelihood is
\begin{equation}\label{eq:lkd}
P(D|g,\mu,\lambda) = \frac{1}{N!} \exp\left(-\int_{[g]} \lambda(z)dz\right)
                     \prod_{a=1}^N \lambda \int_{[g]}\Pr\{M_a=m_a|x_a,g,\mu\} dx_a.
\end{equation}

We compute $\lambda\int_{[g]}\Pr\{O(z)>d|z,g,\mu\}dz$ and
the factors $\lambda\int_{[g]}\Pr\{M_a=m_a|x_a,g,\mu\} dx_a$
using recursions related to the pruning recursion of \cite{felsenstein81}.
We begin with $\lambda\int_{[g]}\Pr\{O(z)>d|z,g,\mu\}dz$. A birth at
a generic point $(\tau,i)$
can be shifted to the child node, $(t_i,i)$,
\[\Pr\{O(\tau,i)>d|(\tau,i),g,\mu\}=\Pr\{O(t_i,i)>d|(t_i,i),g,\mu\}\exp(-\mu(\tau-t_i)),\]
and the integral
over $[g]$ reduced to a sum over contributions from edges:
\begin{equation}
\label{eq:lkdfac1}
\lambda\int_{[g]}\Pr\{O(z)>d|z,g,\mu\}dz = \frac{\lambda}{\mu}\sum_{\bra i,j\ket\in E} \Pr\{O(t_i,i)>d|(t_i,i),g,\mu\}(1-e^{-\mu(t_j-t_i)}).
\end{equation}
We are interested in the cases $d=0$ and $d=1$. Let
$u^{(d)}_i\equiv\Pr\{O(t_i,i)=d|(t_i,i),g,\mu\}$ so
\[
\Pr\{O(t_i,i)>d|(t_i,i),g,\mu\}=\left\{\begin{array}{ll}
  1-u^{(0)}_i & d=0, \\
  1-u^{(0)}_i-u^{(1)}_i & d=1.
\end{array}\right.
\]
We give recursions for the $u^{(d)}_i$.
Consider a pair of edges $\bra j,i\ket, \bra k,i\ket$ in $E$. Let $\delta_{i,j}=e^{-\mu(t_i-t_j)}$.
The recursions
\begin{equation}
 \begin{array}{rcl}
    u^{(0)}_i &=& \left((1-\delta_{i,j})+\delta_{i,j}u^{(0)}_{j}\right) \left((1-\delta_{i,k})+\delta_{i,k}u^{(0)}_{k}\right) \\
    u^{(1)}_i &=& \delta_{i,j}(1-\delta_{i,k})u^{(1)}_{j}+\delta_{i,k}(1-\delta_{i,j})u^{(1)}_{k}+ \delta_{i,j}\delta_{i,k}(u^{(1)}_{j}u^{(0)}_{k}+u^{(0)}_{j}u^{(1)}_{k})
 \end{array} \label{eq:urec1}
 \end{equation}
are evaluated from $u^{(0)}_i=0$ and $u^{(1)}_i=1$ at leaves $i\in V_L$.

We need to compute $\lambda\int_{[g]}\Pr\{M_a=m_a|x_a,g,\mu\}dx_a$
for generic trait patterns.
Trait $c_a$ is born into an edge ancestral to all the leaf nodes which display it,
so the edges of $g$ which contribute to the integral $dx_a$
are those edges, $E_a$ say, on the path to node $R^*$ from the most recent common ancestor
of the leaf nodes in $m_a$. Also, $m_a$ is non-empty, so
$\Pr\{M_a=m_a|(\tau,i),g,\mu\}=\Pr\{M_a=m_a|(t_i,i),g,\mu\}\exp(-\mu(\tau-t_i))$. We write
the integral over $[g]$ in terms of a sum over contributions from edges:
\begin{equation}
  \label{eq:lkdfac2}
  \lambda\int_{[g]}\Pr\{M_a=m_a|x_a,g,\mu\}dx_a = \frac{\lambda}{\mu}\sum_{\bra i,j\ket\in E_a} \Pr\{M_a=m_a|(t_i,i),g,\mu\}(1-e^{-\mu(t_j-t_i)}).
\end{equation}
Let $V_L^{(i)}$ be the set of leaf nodes in $V$ descended from node $i$,
including $i$ if node $i$ is a leaf. For leaf sets $m_a$ let $m_a^{(i)}=V_L^{(i)}\cap m_a$.
Consider two edges $\bra j,i\ket, \bra k,i\ket$ in $E$. Events are independent down the
two branches,
\[
  \Pr\{M^{(i)}_a=m^{(i)}_a|(t_i,i),g,\mu\}=\Pr\{M^{(j)}_a=m_a^{(j)}|(t_i,j),g,\mu\}\Pr\{M^{(k)}_a=m_a^{(k)}|(t_i,k),g,\mu\},
\]
and moving from the top $(t_i,j)$ to the bottom $(t_j,j)$ of branch $\bra j,i\ket$,
\begin{equation}
\label{eq:mrec}
\Pr\{M^{(j)}_a=m_a^{(j)}|(t_i,j),g,\mu\}=\left\{
\begin{array}{ll}
  \delta_{i,j}\times \Pr\{M^{(j)}_a=m_a^{(j)}|(t_{j},j),g,\mu\}  & \mbox{if $m_a^{(j)}\ne\emptyset$,}\\
  (1-\delta_{i,j})+\delta_{i,j}u^{(0)}_{j}         &\mbox{if $m_a^{(j)}=\emptyset$.}
\end{array}\right.
\end{equation}
The recursion is evaluated from the leaves,
\[
\Pr\{M^{(j)}_a=m_a^{(j)}|(t_{j},j),g,\mu\}=\left\{
\begin{array}{ll}
1& \mbox{if $j$ is a leaf and $m^{(j)}_a=\{j\}$,} \\
0&\mbox{if $j$ is a leaf and $m^{(j)}_a=\emptyset$.}
\end{array}\right.
\]
The recursion need not reach the leaves. It can be evaluated from nodes $j$ satisfying
$m^{(j)}_a=\emptyset$, using \Eqn{eq:mrec} since $u^{(0)}_{j}$
is computed for the $\int_{[g]} \lambda(z)dz$ evaluation.

\section{Prior models on trees}
\label{sec:treeprior}

In this section we specify two families of probability distributions over trees,
which we use to represent prior information concerning the phylogeny.

One tree prior we use is a branching process $G_L$ with rate $\theta$
stopped at the instant of the $L$th branching event (counting the branching at the root).
Denote by $\Gamma$ the space of $G_L$-realizable trees and by $dg$ the measure
$\prod_{i\in V_A} dt_{i}$, with counting measure on
topologies. The process $G_L$ determines a density
\[f_G(g|\theta)\propto\theta^{L-1}\exp(-\theta|g|)\]
with respect to $dg$,
where $|g|$ is the sum of all branch lengths, excluding the branch
$\bra R,R^*\ket$. The same functional form of the density is used
when tree leaves are offset in time.

In \Sec{sec:data}, a hypothesis of the form ``$t_R\in[t_{\rm min},t_{\rm max}]$"
is central. This motivates a prior which is non-informative
with respect to such hypotheses.
One prior which is strongly informative for
$t_R$ is the prior density $f(g|T)\propto \mathbb{I}_{t_R\le T}$, the uniform
distribution over all trees in $\Gamma$ with root age smaller than $T$, a fixed upper limit.
We find that, for trees with isochronous leaves at $t_1=t_2=\ldots =t_L=0$, the
marginal distribution of $t_R$ is $t_R^{L-2}$ (for each $g\in\Gamma$ the topology-constrained
volume integral $\int dt_{L+1}\ldots dt_{2L-2}\propto {t_R}^{L-2}$).
This prior represents a state of belief in which
$\Pr\{t_R\in [T/2,T]\}$ is about $2^{L-1}$ times greater than $\Pr\{t_R\in [0,T/2]\}$.
The marginal density of $t_R$ in
the prior \[f_R(g|T)\propto t_R^{2-L}\mathbb{I}_{t_R\le T}\]
is uniform in $[0,T]$.

In Section~\ref{sec:caldata},
certain groups of taxa, called clades, are known to group together on the tree.
Upper and lower bounds on
the age of their common ancestor are used to calibrate rate parameters.
Admissible trees $g\in \Gamma'$, $\Gamma'\subseteq\Gamma$,
satisfy these prior calibration constraints.
Where calibration constraints are imposed, the prior $f_R$ must be modified, in order to maintain
a uniform prior distribution for the root age.
Nodes in clades with clade root times bounded above by calibration constraints
do not contribute a factor $t_R$ to the tree-topology constrained
volume integral $\int \prod_{i\in V_A\setminus \{R\}} dt_{i}$.
The density $f_R$ must be further modified to take into
account non-isochronous leaf dates. The exact result is beyond us.
However, if $S$ is a list of free nodes, $ie$ nodes $i\in V_A\setminus\{R\}$
outside or above root-bounded clades,
and for node $i\in S$ in tree $g\in\Gamma'$,
$s_i$ is the minimum time-value node $i$ can achieve in any admissible tree,
then
\[f_R(g|T)\propto \mathbb{I}_{t_R<T}\prod_{i\in S}(t_R-s_i)^{-1}\]
gives a reasonably flat marginal distribution for $t_R$ large compared to the calibration dates.
Refer to the supplement for the results of prior simulation.
In the examples following Section~\ref{sec:data}, we summarize posterior
distributions computed under tree prior $f=f_R$, with clade
calibration constraints. Results for the prior $f_G$ are similar,
and are displayed in the supplementary material.

We encounter data in which leaf node times
are themselves subject to uncertainty. Calibration data on leaf node times
allow leaf times to vary in a range, so that for each $i\in V_L$, $t_i\in [t^-_i,t^+_i]$. The leaf times
$t_i, i\in V_L$ become missing data. In Section~\ref{sec:data}, the allowed range for leaf times is small compared to the time
over which traits evolve. We take a prior uniform in $[t^-_i,t^+_i]$ for $t_i, i\in V_L$.

\section{Posterior distributions}
\label{sec:cal}

Our final expression for the likelihood is obtained by substituting \Eqn{eq:lkdfac1}
and \Eqn{eq:lkdfac2} into \Eqn{eq:lkd}, and evaluating these terms using
\Eqn{eq:urec1} and \Eqn{eq:mrec} respectively. Multiplying that likelihood by
the tree-prior $f_G$ given in \Sec{sec:treeprior} and a prior density $p(\mu,\lambda,\theta)$ for our rate parameters,
we obtain the posterior distribution
\begin{eqnarray}
p(g,\mu,\lambda,\theta|D)d\theta d\lambda d\mu dg&\propto&\exp\left(-\frac{\lambda}{\mu}\sum_{\bra i,j\ket\in E}
                                        \Pr\{O(t_i,i)>d|(t_i,i),g,\mu\}(1-e^{-\mu(t_j-t_i)})\right) \nonumber\\
     & & \times \quad \prod_{a=1}^N\sum_{\bra i,j\ket\in E_a} \Pr\{M_a=m_a|(t_i,i),g,\mu\}(1-e^{-\mu(t_j-t_i)})\nonumber \\
     & & \times \quad \left(\frac{\lambda}{\mu}\right)^N \theta^{L-1}e^{-\theta|g|} p(\mu,\lambda,\theta) d\theta d\lambda d\mu\prod_{i\in V_A} dt_{i}. \label{eq:post}
\end{eqnarray}
\Eqn{eq:post} holds for tree prior $f_G$. Under tree prior $f_R(g|T)$ we drop parameter $\theta$ from the posterior
and replace $f_G\propto\theta^{L-1}e^{-\theta|g|}$ with $f_R\propto t_R^{2-L}\mathbb{I}_{t_R<T}$.

Time scale is undetermined under scale invariant priors $p(\mu,\lambda,\theta)=(\mu\lambda\theta)^{-1}$. For $\rho>0$, the transformation
$(t_1,\ldots,t_R,\mu,\lambda,\theta)\rightarrow (t_1/\rho,\ldots,t_R/\rho,\mu\rho, \lambda\rho, \theta\rho)$ leaves $p(g,\mu,\lambda,\theta|D)dg d\lambda d\mu d\theta$
invariant, so it cannot be a proper distribution.
The problem remains (for $\rho>t_R/T$) under tree prior $f_R$.
Date calibration data described in \Sec{sec:treeprior} and \Sec{sec:caldata}
restricts the space of tree states from $\Gamma$ to $\Gamma'$,
and thereby breaks the time-rescaling invariance. The posterior becomes proper.

The special case of two taxa (so, $0=t_1\le t_2\le t_3$ with $t_R=t_3$ the tree height)
is of interest for checking and debugging.
The data, $D^{(L)}=(H_1,H_2)$, are two lists of trait instances, including traits present
at just one leaf. \cite{huson04} compute the MLE, $|g|^*$, for the total tree length
$|g|=2t_R-t_2-t_1$ in a two leaf tree directly, using the reversibility of the birth-death process
of traits between the two leaves, and conditioned on $\lambda/\mu$ known.
In this way they motivate a new measure of the distance between
two binary sequences as the pairwise maximum likelihood distance
between the two sequences.
Let $n_1=\card H_1\setminus H_2$, $n_2=\card H_2\setminus H_1$ and $n_{12}=\card (H_1\cap H_2)$,
so that $N=n_{12}+n_1+n_2$. The data $D$ amounts to $n_1,n_2,n_{12}$ in the two leaf case.
The likelihood for the two leaf case, computed from
\Eqn{eq:lkd}, using \Eqn{eq:lkdfac1} and \Eqn{eq:lkdfac2}, is
\[
  P(n_1,n_2,n_{12}||g|,\lambda,\mu)\propto \left(\frac{\lambda}{\mu}\right)^{N}
\exp\left(-\frac{\lambda}{\mu}\left[2-e^{-\mu|g|}\right]\right)
     \left(1-e^{-\mu|g|}\right)^{n_1+n_2}e^{-\mu|g|n_{12}}.
\]
Maximizing this expression over $|g|$ given $\lambda/\mu$ we recover the branch length
calculated in \cite{huson04}. If instead we
maximize $P(n_1,n_2,n_{12}||g|,\lambda,\mu)$ over $\lambda$ and $|g|\ge t_2-t_1$
we get an estimate $|g|^*$ for the time separation of two taxa,
\begin{equation}\label{eq:ml}
|g|^*=\frac{1}{\mu} \log\left(1+\frac{n_1+n_2}{2n_{12}}\right).
\end{equation}
\cite{swadesh52} fits a relation of this kind to lexical trait data.

The posterior distribution for $|g|$ given $\mu$,
which is available in closed form for the two leaf tree,
is useful for debugging MCMC code.
Taking priors $p(\lambda,\theta)=(\lambda\theta)^{-1}$ in \Eqn{eq:post}
and integrating out $\lambda$ and $\theta$, we obtain,
\begin{equation}\label{eq:twopost}
p(|g||\mu,n_1,n_2,n_{12}) \propto \frac{1}{\mu|g|}\left[\frac{e^{-\mu|g|}}{2-e^{-\mu|g|}}\right]^{n_{12}}
\left[\frac{1-e^{-\mu|g|}}{2-e^{-\mu|g|}}\right]^{n_1+n_2}.
\end{equation}
Here $\mu$ and $|g|$ appear in the combination $\mu|g|$. When we consider
large trees, and estimate $\mu$, calibration constraints fixing the age of clades in $g$
separate this pair of variables.

\section{Markov chain Monte Carlo}
\label{sec:mcmc}

We work exclusively with the marginal posterior density $p(g,\mu|D)$. When the
prior for $\lambda$ is $\lambda^{-1}$, this variable is Gamma distributed in the
posterior, and may be integrated. The same observation applies to $\theta$, when we
use the $f_G$ prior.
Sampling the posterior distribution $p(g,\mu|D)$ $via$ Metropolis-Hastings
Markov chain Monte Carlo is fairly straightforward, once efficient schemes for evaluating and updating the recursions,
Equations~(\ref{eq:urec1}) and (\ref{eq:mrec}) have been implemented.

We use the tree operations described in \cite{drummond02}. These include updates which alter the tree topology,
updates which vary node times, updates which vary parameters such as $\mu$, and updates
which make some combination of these changes.
In a specimen update we
generate candidates for Metropolis-Hastings updates
by simulating $\rho\sim U(1/2,2)$ and setting $t'=\rho t$ and $\mu'=\mu/\rho$,
since this is expected to be a ridge direction of the loglikelihood.
In the acceptance probability for this update, the probability density to generate the reverse update,
with $\rho'=1/\rho$, is equal to the probability density to generate the forward update,
and a Jacobian term $|\partial(g',\mu',\rho')/\partial(g,\mu,\rho)|=\rho^{L-2}$ appears in the Hastings ratio.
The calibration constraints fix certain taxa groupings as clades, and bound the
age of the most recent common ancestor of certain clades of taxa. These constraints
are implemented by rejecting proposed states that violate the constraints.

Our MCMC convergence analysis, based on monitoring the asymptotic behavior of the
autocorrelation for $\mu$, $t_R$, and the log-prior and log-likelihood, follows \cite{geyer92a}.
We made a number of checks on our implementation. We check that the computer function
for the likelihood \Eqn{eq:lkd} sums to one over
data. We check that the marginal prior distribution of $t_R$ under $f_R$ with
isochronous leaves is uniform.
We recover the posterior distribution in \Eqn{eq:twopost} in the two leaf case.
We fix a data set and vary the proportions in which update types are used.
We check that statistics computed under the posterior do not vary, to within estimated errors.
We recover the parameters of synthetic data, and the posterior distribution concentrates
on the correct parameter values as the number, $N$, of traits displayed in the data increases.

\section{Data}
\label{sec:data}

\subsection{Word lists}
\label{sec:wordlistdata}

In the \cite{dyen97} and \cite{ringe02} data, a trait is a homology class of words.
The setup is illustrated in Table~\ref{tab:cognates}.
\begin{table}
\caption{\label{tab:cognates} A miniature lexical ``data set''
  with $L=3$ languages, $K=3$ meanings and $N=6$
  distinct traits,
    $C=\{1,2,\ldots,6\}$, from \protect\cite{dyen97}.\vspace*{0.2in}}

    \centering
  \begin{tabular}{r|lll}
      & ``to give''  & ``big'' & ``we''  \\[0.05in]
    \hline
    & & \\
    Flemish  & geven & groot & wy  \\
    Danish & give  & stor & vi \\
    Kashmiri  & dyunu &  bodu  &  asi
  \end{tabular}\qquad $\Longrightarrow$\qquad
  \begin{tabular}{c|ccc}
      & $\mathstrut k=1$ & $k=2$ & $k=3$ \\[0.05in]
    \hline
    & & \\
    $\mathstrut i=1$ & $c=1$ & $c=3$ & $c=6$ \\
    $i=2$ & $c=1$ & $c=4$& $c=6$ \\
    $i=3$ & $c=2$ & $c=5$ & $c=6$
  \end{tabular}
\end{table}
A set of $K$ meaning categories are chosen and, for each of the $L$ languages in the study,
words in the $K$ meaning categories are gathered. The \cite{dyen97} data uses the \cite{swadesh52} ``word list"
(in fact a list of meanings). In this list, $K=200$ core meaning categories
(``All'', ``And'', ``Animal'',\ldots) are given.
Words in the Swadesh meaning
categories are relatively resistant to lateral trait transfer, referred to here as {\it borrowing}.
\cite{embleton86} observes that words borrowed from French and Latin make up about 60\% of
the English lexicon, but less than 6\% of the Swadesh 200-word list. The
\cite{ringe02} data we have uses
a list of $K=328$ meanings, (plus morphological traits, which we do not treat).
There is a Swadesh list of $K=100$ meaning categories thought to be particularly
resistant to borrowing. The word lists are nested, so both data sets
include the 200-word and 100-word lists.

In the following,
trait data collected by \cite{gray03}
for Hittite, Tocharian A and Tocharian B are analysed with 84 languages
(displayed in \Fig{fig:Dfr100-ps}) from the \cite{dyen97} data.
These merged data are referred to hereafter as the
\cite{dyen97} data.
Of the $L=24$ languages in the \cite{ringe02} data (displayed in \Fig{fig:Rfr100-17}), $20$ are ancient.
In contrast, of the $L=87$ languages in the \cite{dyen97} data, just the three
added by \cite{gray03} are ancient. The two data sets are substantially independent.
Both data sets are available in electronic format.

The linguist identifies homology classes among the words in a given  meaning category.
In order to avoid false identification of homology, where there is merely a chance likeness of
sound, linguists require close correspondence of meaning.
Where words are judged to be descended
from a common ancestor they are assigned the same trait label.
This operation, which
requires expert knowledge, is equivalent to
replacing words with trait labels, $c\in C$, and
thereby generating for each
language $i=1,2,\ldots,L$ and each meaning category $k=1,2,\ldots,K$
a trait set $H^{(k)}_i$. In the context of this application,
homology classes of traits are called {\it cognate classes}.
Both data sets mark some cognate classes as equivocal, and offer
``splitting'' and ``lumping'' versions of the data. We present results for the
``splitting'' data which assigns separate labels to cognate classes which may in fact
display a single homologous trait. Results for the lumping data are very similar.
We comment on this systematic error in \Sec{sec:mmsplit}.
\cite{gray03} register the ``splitting" \cite{dyen97} and \cite{ringe02}
cognate data respectively as $87\times 2665$ and $24\times 3174$ binary matrices.

In the example in Table~\ref{tab:cognates},
the data is coded
$H^{1}_1=\{1\}$, $H^{1}_2=\{1\}$, $H^{1}_3=\{2\}$,\ldots,$H^{3}_3=\{6\}$.
Looking at \Sec{sec:model}, we have
an extra superscript $(k)$ on trait-sets $H_i$ marking the meaning class. In \Sec{sec:inference}
we start with one independent copy of the trait birth-death process
$H(\tau,i)$ for each meaning category.

The vocabularies of some ancient languages are only partially reconstructed,
creating gaps in the binary sequence data. The \cite{ringe02} data marks these gaps.
We are unable to treat missing data at this stage. We are obliged to drop
from the analysis of the \cite{ringe02} data the languages
Gothic, Lycian, Luvian, Oscan, Umbrian, Old Prussian, Old Persian,
Avestan and Tocharian A, leaving the languages in \Fig{fig:Rfr100-17}.
We retain some languages with small numbers of gaps, simply marking the gap as trait-absence.
We discuss the associated model mis-specification bias in \Sec{sec:mmzeros}.
The number of gaps in our registration of the \cite{dyen97} data is negligible.

\subsection{Calibration data}
\label{sec:caldata}

Historical sources provide rate calibration data for these Indo-European data sets.
\cite{atkinson05} compile calibration points. For example,
the Brythonic languages
{\it Welsh\verb|_|N, Welsh\verb|_|C, Breton\verb|_|List, Breton\verb|_|SE} and {\it Breton\verb|_|ST}
form a clade in the \cite{dyen97} data, with
a common ancestor between 1450 and 1600 years before the present (BP, where the present is
the year 2000 - only roughly the time the data was gathered, because the dating accuracy is in
any case low).
In our analysis of the two data sets we imposed 16 groups of taxa as clades:
Brythonic, Celtic, Italic, Iberian-French, Germanic, West Germanic, North Germanic,
Balto-Slav, Slav, Indic, Indo-Iranian, Iranian, Albanian, Greek, Armenian and Tocharic.

The calibration points marked by horizontal bars in the sample tree states below
give both lower and upper bounds on clade root times. Each such calibration point
gives an independent estimate for $\lambda$, $\mu$ and $\theta$.
Prior knowledge providing only a lower bound on language branching (``languages
A and B were distinct by year C'') is more common, but less valuable,
as it does not break the scale invariance discussed in \Sec{sec:cal}.
\cite{yang06} observe that, in a phylogenetic setting, there is often
good evidence for the lower bound, but little confidence in the upper bound.
The same applies here, since the upper bound for a split is supported by
the absence of historical evidence for separate vocabularies.
However, uncertainties in the positions of the
upper bounds are of the order of hundreds of years, whilst the evolution rates
we are calibrating give half lives of thousands of years, so we have not
pursued this source of uncertainty.
We do see one result which suggests that a soft upper limit would
            have improved the analysis. The one incorrect clade age estimate (Balto-Slav) we see
            in the cross-validation study (Supplement, Section 9.4) fell
            above the upper limit of the calibration interval. That upper limit
            is different from the others, since it is not based upon historical texts,
            but is instead derived from consideration of excavated cultural remains.
            The link between language and excavated culture is obviously weaker
            than language and literature.

\subsection{Previous studies}
\label{sec:previous}

The survey given in \cite{sankoff73} summarizes models of cognate
trait data. \cite{sankoff73} presents relatively realistic models
which are complex and parameter-rich. \cite{sankoff73} discusses
inference based on pairwise distances between the
binary-trait data-vectors of two languages. This mode of inference
has been the norm for lexical cognate class data. Thus \cite{dyen92} use
classical hierarchical clustering of data-vectors based on pairwise
distances between languages to establish a tree of languages.
In contrast, \cite{gray03} use \cite{ronquist03} MrBayes software and the Bayesian
phylogenetic methods of \cite{yang97}, to fit the finite-sites DNA
sequence model of \cite{felsenstein81}. The MrBayes software allowed them to
account for the thinning of traits surviving into zero taxa.
\cite{pagel06} describe and fit a related, more realistic,
model of cognate replacement within
meaning category. These models allow traits identified in the data
as homologous to arise by independent innovation. \cite{warnow06}
propose a model in which each homology class has a unique birth
event. However, there is to date no statistical inference for the
model. \cite{ringe02}, \cite{erdem05} and \cite{nakhleh05} reject
dating, and avoid explicit modelling. They make a parsimony analysis
without explicit measures of uncertainty. They allow some lateral
transfer of traits, and thereby generalize to graphs which are not
trees. They employ expert linguistic intervention in the inference,
which becomes a well informed search through phylogenies. In light
of the random and systematic error we measure below, we do not
expect estimators of tree topology related to the mode ($ie$
parsimony) to be adequate. However \cite{ringe02} add morphological
traits. These may be more reliable data than cognate traits. Such
traits can be analysed in the framework we set out. \cite{garrett06}
shows that the breakup of dialect continua into languages is not tree-like.
The local borrowing model we give in \Sec{sec:mmborrow} generates
similar model violations.

\section{Inference}
\label{sec:inference}

\subsection{Fitted Models}
\label{sec:fitted}

When we fit the model of \Sec{sec:model} to the \cite{dyen97} and \cite{ringe02} data,
we identify a model mis-specification problem.
For meaning classes $k=1,2,\ldots,K$ denote by $H^{(k)}(\tau,i)$ a trait birth-death
process modelling the evolution of words in meaning category $k$, so that
for $i=1,2,\ldots,L$, $H^{(k)}_i=H^{(k)}(t_i,i)$ is the data at the leaves under NOABSENT.
Let $\lambda^{(k)}$ and $\mu^{(k)}$ be the birth and death rates for traits in meaning class $k$.
It is reasonable to expect any real language to have at least one word in each of the
semantic fields in the Swadesh 200-word list at all times. It follows that
the birth-death process must satisfy a {\it no-empty-field} condition, $H^{(k)}(\tau,i)\ne\emptyset$
or $N(\tau,i)>0$, for each $(\tau,i)\in [g]$.

We ignore this no-empty-field condition in our analysis.
We lump together the $K$ copies of
the birth-death process of traits corresponding to the different meaning classes.
Under the empty-field approximation, and assuming the death rates
$\mu=\mu^{(k)},\ k=1,2,\ldots,K$ are all equal (see \Sec{sec:mmratestraits}), the superposition
\begin{equation}\label{eq:merge}
    H(\tau,i)=\bigcup_{k=1}^K H^{(k)}(\tau,i)
\end{equation}
of birth-death processes generates another instance of the same process, with birth rate
$\lambda=\sum_k \lambda^{(k)}$
and death rate $\mu$. If the mean number of words per meaning category is large,
then the process does not visit the constraint, so the approximation holds.
As we show in \Sec{sec:mmefa}, this condition is not satisfied. However, when
we study synthetic data simulated under the empty-empty-field condition,
the calibration constraint forces the fitting procedure to adapt the approximating model
to data by distorting estimates of $\mu$, and thereby reproduce the uncalibrated
clade root ages very well.

We carry out MCMC from posterior distributions $p(g,\mu|D)$ determined by \Eqn{eq:post} and
the \cite{dyen97} and \cite{ringe02} data, under the NOUNIQUE observation model.
We repeated the analysis with NOABSENT for the \cite{ringe02} data, obtaining similar results.
We apply the branching process prior $f_G$
with prior $1/\theta$, and the uniform root prior $f_R$ with $T=16000$ (an uncontroversial upper limit on $t_R$).
The data overwhelm these two priors, differences between posterior estimates obtained under
the two priors are slight, and we therefore discuss results for the prior $f_R$ in this paper
and very briefly report $f_G$-results in the supplement.
Results are completely insensitive to the choice of $T$, for all $T$ sufficiently large.

In our search for conflicting signals in the data, we analyzed (in addition) subsets of the data.
As discussed in Section~\ref{sec:mmborrow}, analyses of subsets of languages may
be less exposed to error due to certain forms of borrowing. On the other hand
we may uncover rate heterogeneity between word lists or between groups of languages.
We reduce the \cite{dyen97} data to the Swadesh 100-word list, and the \cite{ringe02} data
to the Swadesh 200- and 100-word lists. We thin the \cite{dyen97} data from $L=87$ languages
down to two sets containing $L=31$ languages, and $L=30$ languages (the two subsets are displayed
in the supplementary material),
chosen in such a way that the pivotal calibrating dates remain applicable.
These two data subsets overlap at 8 languages, but just one of the five calibration points
has any common data (Tocharian, where there is no choice).
We label analyses ``$\mbox{\rm Data-1st-author} / {\rm Word\ List}/ {\rm Number\ of\ Leaves}$''.

\subsection{Results}
\label{sec:results}

\begin{figure}[htbp]
  \includegraphics[width=5.5in]{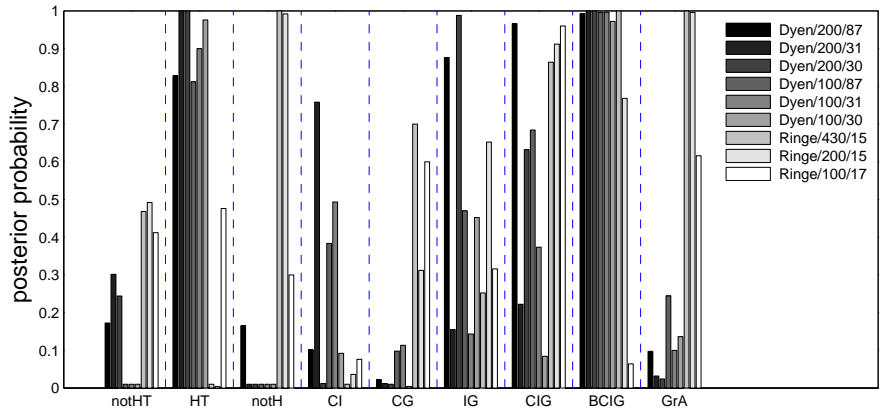}\\
  \caption{Posterior probabilites for selected clades, across data sets.
  $x$-axis labels: {\sf BGCI}, Balto-Slav-Germanic-Celtic-Italic; {\sf GCI}, Germanic-Celtic-Italic; {\sf CI}, Celtic-Italic;
  {\sf CG}, Celtic-Germanic; {\sf GI}, Germanic-Italic; {\sf GrA}, Greek-Armenian;
  $\sf notHT$, complement of Hittite-Tocharian; $\sf notH$, complement of
  Hittite. }\label{fig:cladeprob}
\end{figure}

\begin{figure}[htbp]
  \includegraphics[width=5.5in]{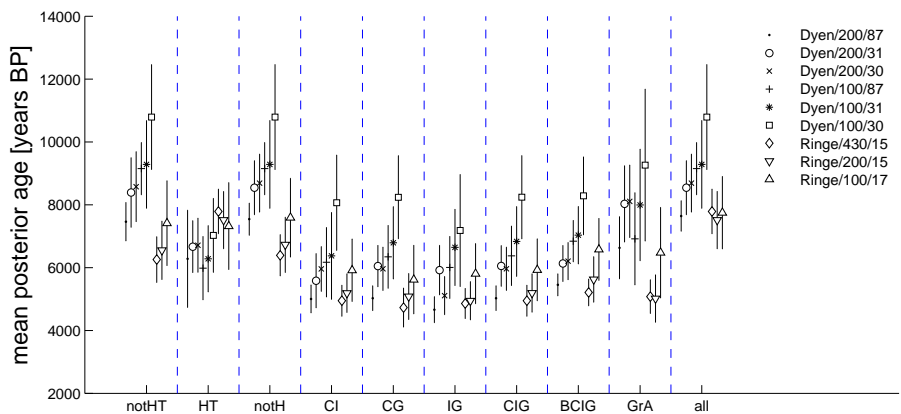}\\
  \caption{The mean posterior ages, in years BP, for
  the most recent common ancestor (MRCA, super-clade root)
  of languages in selected combinations of clades ($ie$ super-clades).
  $x$-axis labels as for \protect\Fig{fig:cladeprob}. }\label{fig:cladeage}
\end{figure}

Figures~\ref{fig:cladeprob} and \ref{fig:cladeage}
give a compact quantitative summary of the ${\rm Dyen}/200/87$, ${\rm Dyen}/100/87$,
${\rm Dyen}/200/31$, ${\rm Dyen}/100/31$, ${\rm Dyen}/200/30$, ${\rm Dyen}/100/30$
${\rm Ringe}/328/15$, ${\rm Ringe}/200/15$ and ${\rm Ringe}/100/17$ posterior distributions.
The posterior probabilities for a selection of clades are displayed in \Fig{fig:cladeprob},
and clade labels {\sf BGCI}, {\sf GCI}, {\sf CI},  {\sf CG},  {\sf GI},  {\sf GrA}, $\sf notHT$ and $\sf notH$
defined.
The posterior mean age for the common ancestor of the languages defining each corresponding clade
is displayed in \Fig{fig:cladeage}. This format is useful for identifying conflict
between data subsets, once clades of interest have been identified.

\begin{figure}[htbp]
  \includegraphics[height=8in]{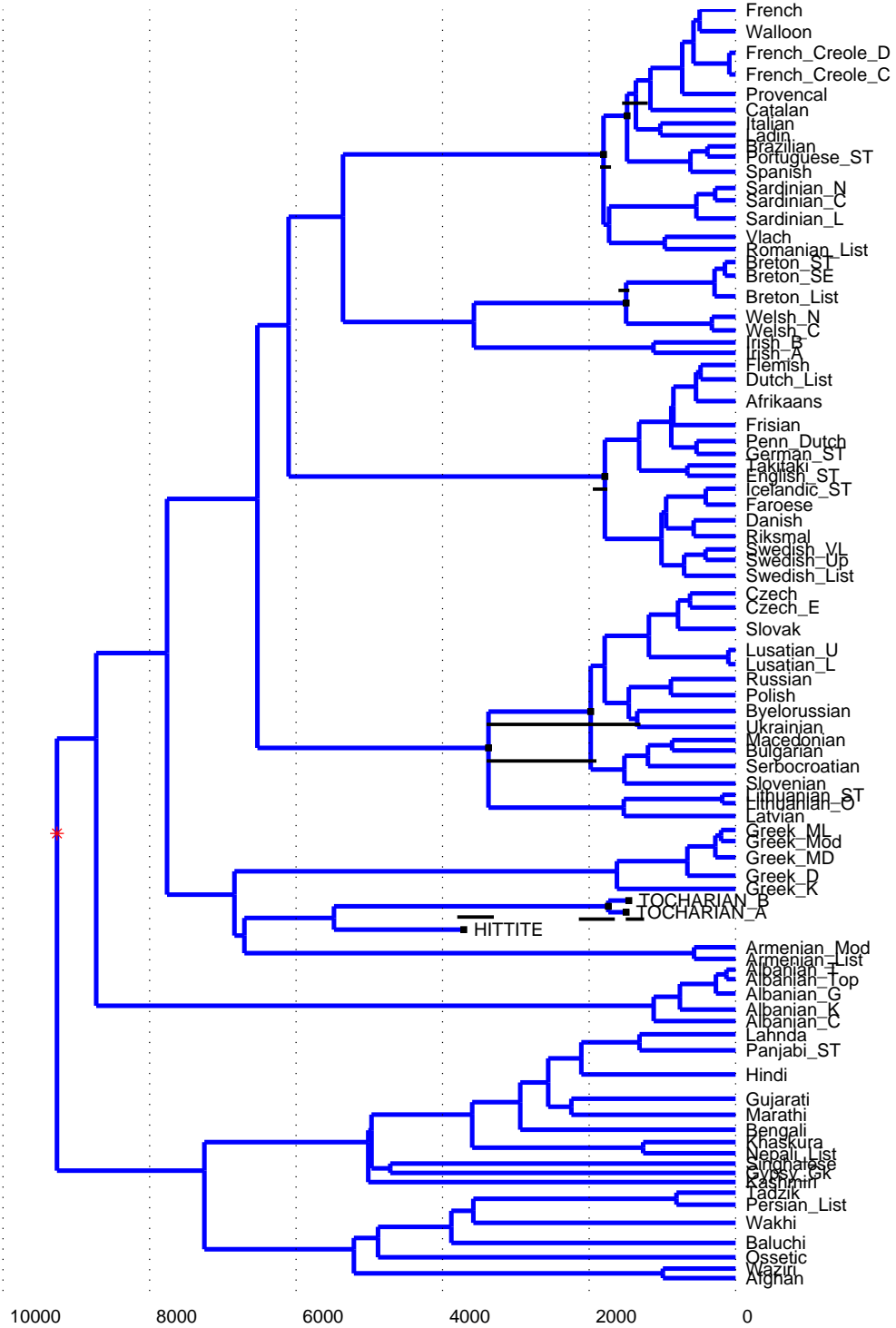}
  \caption{Tree sampled from the ${\rm Dyen}/100/87$ posterior distribution.
  $x$-axis gives age in years.
  Prior constraints on eight clade root and three leaf ages are indicated
  by horizontal bars. In order to reduce clutter, a single bar shows the two Tocharian A and B constraints,
  which are near equal.
  }
  \label{fig:Dfr100-ps}
\end{figure}

\begin{figure}[h]
  \[\includegraphics[width=4.5in]{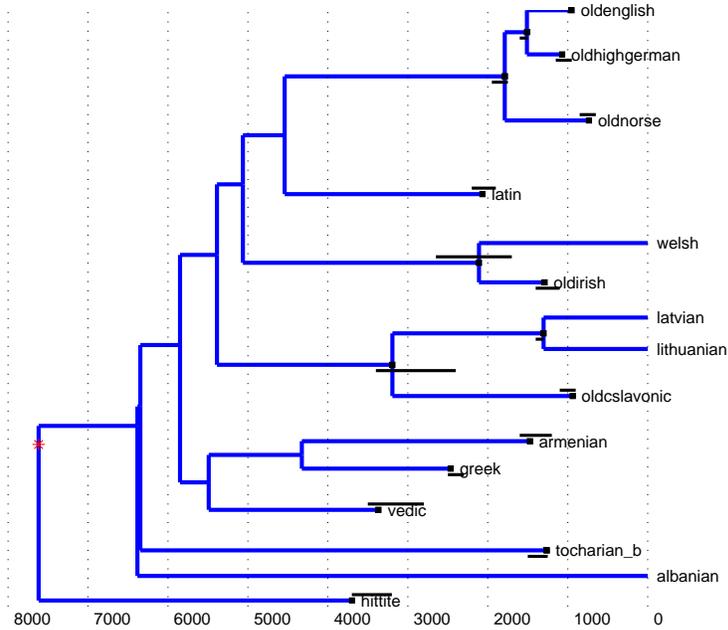}\]
  \vspace*{-0in}
  \caption{
  A sampled state
  illustrating the ${\rm Ringe}/100/17$ posterior distribution.
  Prior constraints on topology impose 7 clades. Prior uncertainties in clade root and leaf ages are indicated
  by horizontal bars.}\label{fig:Rfr100-17}
\end{figure}

In the supplement, \cite{nicholls07}, we define and display consensus trees,
a central point estimate
for topology and branch length.
The consensus tree is not a state in the sample space of trees. Prior constraints and leaf ages
are represented on sampled states so we give, in Figures~\ref{fig:Dfr100-ps}
and \ref{fig:Rfr100-17}, samples drawn from the ${\rm Dyen}/100/87$ and ${\rm Ringe}/100/17$
posterior distributions. 

Referring to \Fig{fig:cladeprob}, there is some conflict in the support
for clades across analyses. The {\sf BCIG}
group is strongly supported in all analyses (except ${\rm Ringe}/100/17$, which does at least allow it).
The {\sf CIG} group is supported in all analyses (except ${\rm Dyen}/100/30$, which allows it).
However all the sub-clades {\sf CI}, {\sf CG} and {\sf IG} are at odds with at
least one data set.
Age estimates for the common ancestor of the languages in the {\sf CIG} clade are, in all analyses,
close to the age estimates for the common ancestors of the subclades, suggesting the breakup
occurred in a relatively small interval of time, so the split structure is poorly resolved.

Referring again to the clade probabilities, \Fig{fig:cladeprob},
and the consensus trees in the supplement, Hittite and Tocharian form an outgroup in the three
${\rm Dyen}/200/Y$ analyses, are grouped with Greek and Armenian in the
${\rm Dyen}/100/Y$ analyses, and are split in the three ${\rm Ringe}/X/Y$ analyses.
There are many model mispecification issues for Hittite and Tocharian.
Comparing {\sf notH} and {\sf all}
in \Fig{fig:cladeage}, Hittite adds 1000 years to the posterior mean root age in
the ${\rm Ringe}/328/15$ analysis.
The contrast between the \cite{dyen97} and \cite{ringe02}
analyses is most clearly visible in the {\sf notH} and {\sf notHT}
columns of Figures~\ref{fig:cladeprob} and \ref{fig:cladeage}.

In contrast, conflicts between analyses of the
Swadesh 100-word list (${\rm Dyen}/100/87$, ${\rm Dyen}/100/31$,
${\rm Dyen}/100/30$ and ${\rm Ringe}/100/17$, symbols $+*\Box\triangle$)
are almost absent. Both {\sf CI} and {\sf IG}
(see \Fig{fig:cladeprob}) are allowed by these analyses. ${\rm Ringe}/100/17$ allows
the clade {\sf HT} and does not impose {\sf HT} as an outgroup (which would be a conflict, as
{\sf notHT} is not a clade of ${\rm Dyen}/100/Y$). In other areas of conflict,
the ${\rm Dyen}/100/Y$ analyses allow a {\sf GrA} clade. This lack of conflict
comes at the price of greater random error (compared to analyses on longer word-lists).
One striking conflict remains: the position of Indo-Iranian relative to the root is
quite different in the ${\rm Ringe}/100/17$, and ${\rm Dyen}/100/Y$ analyses.

The posterior mean ages for
the {\sf notH}, {\sf notHT}, {\sf all} and {\sf GrA}, which show particular conflict in
\Fig{fig:cladeage},
are in fair agreement for analyses based on the Swadesh 100-word list.
Comparing the ${\rm Dyen}/100/30$ and ${\rm Ringe}/100/17$ analyses, and looking at
\Fig{fig:cladeage} and Supplement-Fig.~7, the clade root ages for
{\sf notHT} do not agree in either simulation. Otherwise there is
agreement between the four analyses, on the ten measured ages, under one or both
prior weightings. This reduced ($K=100$) set of traits is chosen
to be resistent to borrowing. Posterior predictive replicates computed in \Sec{sec:mmratestraits}
show little evidence of rate heterogeneity within this class of traits. The corresponding words
are relatively well attested in otherwise incompletely
reconstructed ancient languages, so there is little missing data. In \Sec{sec:predictive} we
compute posterior predictive distributions for singleton traits in the ${\rm Ringe}/100/17$ analysis;
these agree well with external data.

In summary, the systematic errors displayed in our four age estimates
from the Swadesh 100-word list are representative.
On the other hand, most features of tree topology
which were in doubt, remain in doubt.


\section{Model mis-specification}
\label{sec:mm}

\begin{figure}[htbp]
  \includegraphics[width=5in]{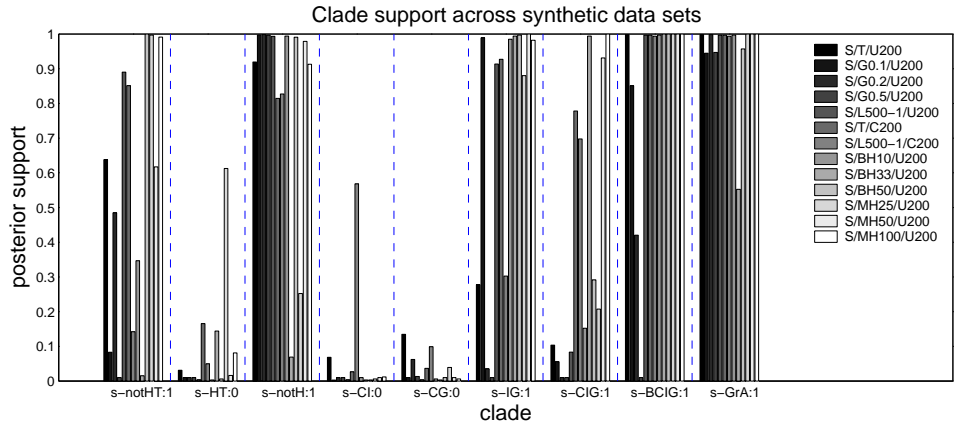}\\[-0.3in]
  \caption{Synthetic data yields estimates of posterior probabilities for selected clades,
  across synthetic data sets.
  $x$-axis labels as for \protect\Fig{fig:cladeprob} with {\sf s-}
  prefix indicating synthetic and 0/1 indicating absence/presence of the clade in the true tree.}\label{fig:synthcladesupport}
  \vspace*{0.1in}
\end{figure}

\begin{figure}[htbp]
  \includegraphics[width=5in]{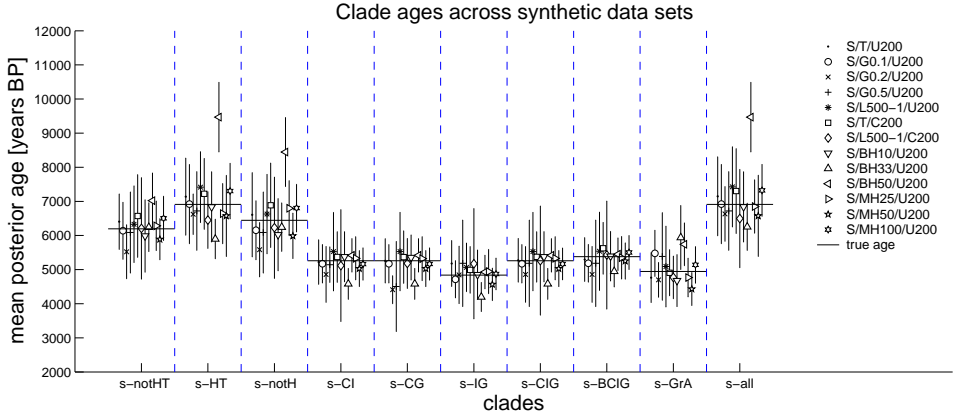}\\[-0.3in]
  \caption{Synthetic data yields estimates of
  mean posterior ages, in years BP, for
  the most recent common ancestor (MRCA, super-clade root)
  as in \protect\Fig{fig:cladeprob} with {\sf s-} prefix indicating synthetic. }\label{fig:synthcladeages}
\end{figure}

We head this section with a summary of its results. These results coincide with
the conclusions we draw from the between-data analyses in \Sec{sec:results}:
our age estimates are robust; tree topology less so.
In \Fig{fig:synthcladesupport} and \Fig{fig:synthcladeages}
we present results from synthetic data,
simulated on a tree sampled from the posterior distribution of the $Ringe/200/15$ analysis
(true clade structure is marked in \Fig{fig:synthcladesupport} with $0$ below false clades
and $1$ below true; the true tree is in \cite{nicholls07}),
under a range of observation models intended to mimic likely model mis-specification. Details of these models,
which simulate the empty-field-condition, plausible levels of borrowing
and branch-wise and trait-wise rate heterogeneity, are given in Sections~\ref{sec:mmborrow} through \ref{sec:mmzeros}.

Clades imposed in reconstructions from synthetic data
are indexed {\sf s-} and are the same as the clades defined below \Fig{fig:cladeprob}
and displayed as horizontal bars in \Fig{fig:Rfr100-17}.
Analyses of synthetic data are indexed ``S/X/Y''. Values of X indicate borrowing and rate heterogeneity:
X=``T'' is no borrowing; X=``G$b$'' is global borrowing at rate
$b\mu$; X=``L$z-b$'' is local borrowing, between languages with a
common ancestor not more than $z$ years in the past, at rate
$b\mu$; and X=``BH$\rho$'' and X=``MH$\rho$'' have rates drawn independently, for each
branch (BH) and meaning category (MH) respectively, from a Gamma distribution with mean $\mu$ and standard
deviation $(\rho/100)\mu$. Values of Y show the constraint applied:
Y=``U$n$'' is the unconstrained birth death process of set
elements, with $n=\lambda/\mu$ the expected number of distinct
traits at each leaf, under the NOABSENT observation model;
Y=``C$n$'' simulates cognate classes under the no-empty-field
constraint, using $n$ meaning categories.


Systematic errors in tree {\it node ages} inferred from synthetic data generated under these models
are in general small. With the exception of $S/BH50/U200$ (see \Sec{sec:mmratesst}),
the systematic error we generated in \Fig{fig:synthcladeages} is of the same order of magnitude
as, or smaller than, random error. Systematic error in estimated {\it rates} $\mu$ (not shown)
is highly significant. Calibration data fixes dates and topology for that part of the
tree adjacent to the leaves, forcing the
inference to accommodate model mis-specification by adjusting rates. The modified rates fit
the imposed trait evolution adjacent to the leaves. If model mis-specification
is homogeneous over the tree, as is the case for the empty-field-approximation,
trait evolution deep in the tree may be well represented by these biased rates,
and date estimates are accordingly robust.

Tree topology is not robust to the model mis-specification we explored.
The ``true'' s-clades {\sf s-GrA},
{\sf s-BCIG}, and the rooting clades {\sf s-notHT}, and {\sf s-notH}
have robust support at levels which do not allow rejection of the truth.
The ``true'' s-clades {\sf s-CIG} and {\sf s-IG} are not well reconstructed
when borrowing is substantial. The branching
at the top of the superclade {\sf s-BCIG}
is poorly resolved as the {\sf s-BCIG}, {\sf s-CIG} and {\sf s-IG} branches are separated by
just 1000 years in the tree on which the synthetic data was simulated (see \cite{nicholls07}),
which is small compared to $\mu^{-1}\simeq 3000$. Nevertheless, the truth is
rejected only at very high levels of borrowing (S/Gb/Y where $b=0.2,0.5$).
Clade age estimates, shown in \Fig{fig:cladeage} and \Fig{fig:synthcladeages},
can be stable across analyses when topology is uncertain. This is because the
super-clade ages are determined largely by total tree length; total
tree length is tightly coupled to the number of transitions on the tree,
which is rather well determined by data.

\subsection{Global and local borrowing}
\label{sec:mmborrow}

Word borrowing from languages outside
the study is straightforward trait birth (unless the same word is borrowed into
several languages). If we delete a source language from our study, we
thereby remove the model error associated with borrowing from that language.
The consistency we see in \Fig{fig:cladeage}
between clade ages reconstructed for near-disjoint
subsets of languages, and the full set, in the ${\rm Dyen}/200/87$, ${\rm Dyen}/200/31$
and ${\rm Dyen}/200/30$ posteriors, suggests that borrowing is not
distorting the ${\rm Dyen}/200/87$ estimates themselves.

Our models of borrowing are as follows.
We associate with each time slice $\tau\in[0,\infty)$ across the tree
a linkage graph $({\cal E}(\tau),{\cal V}(\tau))$ with nodes,
${\cal V}(\tau)=\{(\tau,i); (\tau,i)\in [g]\}$,
corresponding to points in $[g]$ intersected by the slice.
The linkage graph models traffic between languages;
its edges $\bra y,z\ket\in {\cal E}(\tau)$ connect sets $H(y)$ and $H(z)$
between which trait instances can pass.
Let $\tilde {\cal V}(z)=\{z\in {\cal V}(\tau): \exists y\in {\cal V}(\tau), \bra y,z\ket\in {\cal E}(\tau)\}$
be the set of nodes adjacent
to $z\in {\cal V}(\tau)$.
Let $b$ denote the relative rate of word-borrowing to word-death. At per capita rate $b\mu$
each instance of each trait in each language in the time slice  $\tau$
generates a borrowing event. Suppose the selected trait-instance
is in language $z\in{\cal V}(\tau)$ and is labeled $c\in C$.
A language $y\in \tilde {\cal V}(z)$ is chosen,
uniformly at random from nodes adjacent to $z$ on $({\cal E}(\tau),{\cal V}(\tau))$,
and we set $H(y)\leftarrow H(y)\cup \{c\}$, $ie$ the word is copied into
the target language.

We model local borrowing as follows. Words transfer between languages
which have a sufficiently recent common ancestor.
The linkage graph at time $t$ includes an edge from $(t,i)$ to $(t,j)$
if points $(t,i)$ and $(t,j)$ in $[g]$ have a common ancestor
less than $z$ years in the past. In this model linked groups of languages break up into
linked subgroups.
In our model of widespread borrowing (the ``global" borrowing model), all languages communicate equally
with all other languages, and the linkage graph is the complete graph.

Our exploration of these models is summarized
in Figures~\ref{fig:synthcladesupport} and \ref{fig:synthcladeages}
by the three S/G$b$/U200 data sets and the S/L$500-1$/U200 data set.
We display global borrowing at relative rates of 10\%, 20\% and 50\% the death rate.
Higher global rates are probably irrelevant. In the data,
the distribution of $\card(M_a)$, the number of languages displaying cognate $a$,
tails off rapidly, so that few cognates are displayed in many languages.
At $b\simeq 1$, cognates simply survive too well, and many cognates from
deep in the tree survive into many languages.
Local borrowing has time depth $z=500$ and a borrowing rate equal to the death rate.
We see from \Fig{fig:synthcladeages} that age estimates are robust to this form
of model mis-specification.

\subsection{Predictive distributions and external data}
\label{sec:predictive}

Where the observation model is NOABSENT,
singleton traits {\it are} present, and we can use them
to test the model. We drop them from the data,
carry out the inference under NOUNIQUE, and then see if we
can predict the number of singleton traits for each taxon.
This check was available for the \cite{ringe02} data. We expect rate heterogeneity
and borrowing to be visible (but probably not distinguishable)
in these tests. 

Denote by $\tilde D$ synthetic trait data generated under the NOABSENT observation model,
displaying $\tilde N$ distinct traits. For trait $a=1,2,\ldots,\tilde N$ let $\tilde M_a$ give
the indices of leaves displaying an instance of trait $a$ for predicted data $\tilde D$,
and $\tilde X_i$ be the number of singleton traits in $\tilde D$ at taxon $i$,
\[
\tilde X_i=\card \{\tilde M_a: \tilde M_a=\{i\}, a=1,2,\ldots,\tilde N\}\quad i\in V_L.
\]
The posterior predictive distribution $\Pr\{\tilde D|D\}$ is
\[
\Pr\{\tilde D|D\}=\int \Pr\{\tilde D|g,\mu,\lambda\}p(g,\mu,\lambda|D)dg d\mu d\lambda
\]
and this determines a predictive distribution for $\tilde X_i$. We sample
$\mu,\lambda$ and $g$ from the posterior $p(g,\mu,\lambda|D)$ ($g$
and $\mu$ are available from MCMC output; we restore $\lambda$ by sampling
its posterior conditional density), simulate
synthetic data $\tilde D$ at the leaves of $g$, and compute $\tilde X_i$
from $\tilde D$.
Let $X_i(D)=\card \{M_a: M_a=\{i\}, a=1,2,\ldots,N\},$ denote the number
of singleton traits at taxon $i$ in the original real data itself.

Predictive distributions for $X_i$ are given in Supplement-Fig.~9.
The predictive distributions for $\tilde X_i$ over-estimate the $X_i$
in the \cite{ringe02} data with $K=328$ meaning categories.
Since borrowing depletes singleton traits,
this is consistent with model mis-specification due to borrowing.
Also, we expect borrowing to be weaker on the shorter word-lists ($K=100,200$),
since the shorter lists are by design more resistent to borrowing.
We see in Supplement-Fig.~9 that singleton traits are indeed
more reliably predicted on shorter lists (especially $K=100$).
Rate heterogeneity can mimic this behavior.
Corresponding studies for synthetic data are given in Supplement-Fig.~10,11.
Predictive distributions from the shorter word-lists are in good agreement with the
data.

\subsection{Rate heterogeneity across traits}
\label{sec:mmratestraits}

\cite{pagel06} show that the evolution rates of words are, for a given
meaning category, fairly consistent across data sets, whilst varying more substantially between meaning categories.
\Fig{fig:showrates} displays a tendency for the shorter word-lists to evolve at relatively slower rates.
This is expected. However the rate variation between data sets
in \Fig{fig:showrates} does not lead directly to
variation in estimated root times in \Fig{fig:cladeage}. For example, ${\rm Dyen}/200/87$
and ${\rm Dyen}/100/87$ differ by a factor 1.5 in posterior mean rate,
but by just 1.2 in root age.
Time depth measurements do not depend on an assumption of constant rates
{\it between} analyses, since rates are estimated from calibration points in the recent history
of the same data used to predict branching times.

In order to generate synthetic data with rate heterogeneity across meaning classes (the $S/MH\rho/U200$ simulations),
we draw rates \[\mu^{(k)}\sim{\rm Gamma}(\alpha,\beta)\]
independently for each meaning category $k=1,2,\ldots,K$,
with mean $\alpha\beta=\mu$ and variance $\alpha\beta^2=(r\mu)^2$,
where $r=0.25,0.5$ and $r=1$, simulate a trait process $H^{(k)}(\tau,i)$ at rate $\lambda, \mu^{(k)}$,
merge meaning categories, as \Eqn{eq:merge}, and then read off data, as \Eqn{eq:nounique}. \cite{pagel06}
estimate that the rate variance over meaning classes is $(r\mu)^2\simeq \mu^2/9$.

Rate heterogeneity across traits distorts the distribution of the number, $\card M_a$, of
languages in which trait $a\in C$ appears. Denote by $Y^{(n)}=Y^{(n)}(D)$ the number of traits displayed
at $n$ leaves,
\[
Y^{(n)}(D)=\card \{M_a: \card M_a=n, a=1,2,\ldots,N\},
\]
and let $\tilde Y^{(n)}=Y^{(n)}(\tilde D)$ be the corresponding random variable computed from posterior predictive
data $\tilde D\sim \Pr\{\tilde D|D\}$. We plot ${\sf E}(\tilde Y^{(n)}|D)-Y^{(n)}$
and the envelope $\pm 2{\rm std}(\tilde Y^{(n)}|D)$.
In the supplementary material (Supplement-Fig.~12)
we show that the fitting procedure is unable to reproduce
the trait frequency distribution in synthetic data with high levels of rate heterogeneity
across meaning classes (standard deviation $50\%$ of the mean)
but lower levels ($25\%$) are invisible.

Returning to the real data, in Supplement-Fig.~13 (left),
some inconsistency attributable to rate heterogeneity between traits
is visible in our ${\rm Ringe}/328/15$ analysis.
Among other problems, the data contains an excess of traits appearing in 10 or more leaves.
This is caused by a small cohort of traits evolving at death rate $\mu$ small compared
to the rest.
The effect is very greatly reduced in the ${\rm Ringe}/100/17$ analysis
(Supplement-Fig.~13, right).
Analyses of the \cite{dyen97} data show a similar pattern.

\subsection{Rate heterogeneity in space and time}
\label{sec:mmratesst}

\begin{figure}[htbp]
  \vspace*{0.1in}
  \includegraphics[width=5in]{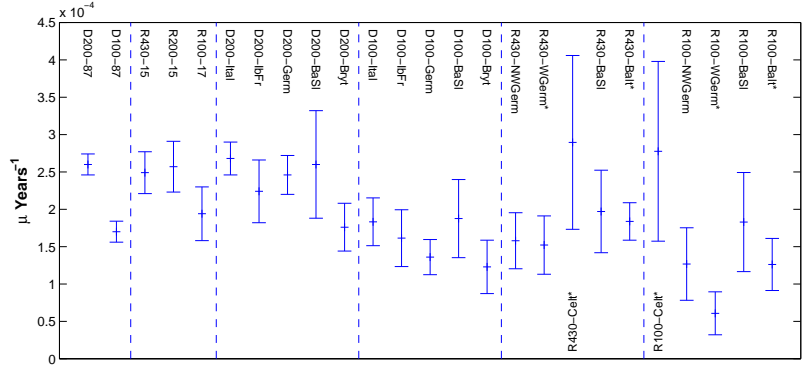}\\[-0.2in]
  \caption{Posterior mean values of $\mu$, with standard errors at
  two standard deviations,
  measured in analyses ${\rm Dyen}/200/87$ (D200-87), ${\rm Dyen}/100/87$ (D100-87),
  ${\rm Ringe}/328/15$ (R328-15), ${\rm Ringe}/200/15$ (R200-15) and ${\rm Dyen}/100/17$ (R100-17),
  and independently using calibration points in distinct clades. The observation model is NOUNIQUE, except
  for two-leaf clades marked *, where NOABSENT must be used.}\label{fig:showrates}
\end{figure}

Time depth measurements depend on
some assumption about the way rates have changed over time {\it within} each data set we analyze.
The variations in rates between clades within each of the D100, D200, R328 and R100 groups in
\Fig{fig:showrates} give us an indication of the un-modelled rate variation we can
expect in the deeper branches of the tree.
We give
the per-trait-instance death rate $\mu$ for each
clade calibration constraint independently. We sampled the posterior distributions $p(g,\mu|D_{\rm clade})$
determined by the data for each clade in turn. We could use the posterior rate distribution from one calibration clade
as a prior to predict the age range for the root of another calibration clade.
Where confidence intervals for the reconstructed rates of a given data set overlap,
the corresponding predictions will be good. Such prediction is legitimate
within one data set only, so we compare rates between vertical dashed lines.

In the supplement for this section we report and discuss a similar
cross-validation exercise on the ${\rm Dyen}/100/87$ analysis.
We drop each calibration-clade in turn (both topology and age constraint)
and estimate the clade root
age using all the word-list data and the remaining calibration points.
Of 10 such tests, 8 succeed. The predicted age range for Balto-Slav is slightly too
deep, and that of Hittite very significantly too young.

Synthetic data with spatio-temporal rate variation ($S/BH\rho/U200$ analyses),
have rates \[\mu_{\langle i,j\rangle}\sim{\rm Gamma}(\alpha,\beta)\]
drawn independently on each edge $\langle i,j\rangle\in E$,
with mean $\alpha\beta=\mu$ and variance $\alpha\beta^2=(r\mu)^2$,
where $r=\rho/100$, so that the standard deviation of the rates is 10\%, 33\% and 50\% the mean rate.
\cite{lees53} sees 20\% variation in rate estimates from pairs of languages.
Results are robust to moderate levels of unstructured, random rate variation of this kind.

Results are of course not robust to structured rate variation, in which, for example, rates
on edges at ages greater than any calibration point are all
larger than any rates in the calibration zone, or a single-taxon outgroup has an extreme rate.
In $S/BH50/U200$, $\sf s-hittite$ happens to have a high rate. Its root is pushed to great age,
and with it goes the root of the tree. The analysis is exposed to rare ``catastrophic''
trait-evolution events outside the calibration zone. We checked that
that no single language or small outgroup is determining the root age
in the ${\rm Ringe}/100/17$, and ${\rm Dyen}/100/Y$ analyses.
Agreement between
reconstructions based on the predominantly ancient languages of \cite{ringe02}
and modern languages of \cite{dyen97} shows that there is
at least no such structured rate variation in the recent past.


%

\subsection{The empty-field approximation}
\label{sec:mmefa}

Our empty-field approximation will be good
if there is significant ``polymorphism'', that is, if the mean number $\lambda^{(k)}/\mu^{(k)}$ of traits ($ie$, words per meaning category)
in the $H^{(k)}(\tau,i)$-process is large.
We estimate $\lambda/\mu$ at $273(9)$ for the \cite{dyen97} Swadesh-200 data and $280(25)$
for the \cite{ringe02} Swadesh-200 data
(posterior standard deviation in parenthesis) and hence $\lambda^{(k)}/\mu^{(k)}\simeq 1.4$.
The probability, $\exp(-\lambda^{(k)}/\mu^{(k)})\simeq 1/4$, to find the unconstrained trait-set process
$H^{(k)}(\tau,i)$ in the empty set at any single fixed point $(\tau,i)\in [g]$ is high enough to cause concern.

We simulate synthetic data from the trait birth-death process constrained to respect the
no-empty-field condition. For each of the $k=1,2,\ldots,K$ meaning classes,
we simulate $N^{(k)}(t_R,R)$ from a Poisson distribution constrained
to be greater than zero, then simulate $H^{(k)}(\tau,i)|N^{(k)}(\tau,i)>0$ in $[g]$. The total rate for
the exponential waiting time to the next event does not include $\mu$ if $N^{(k)}(\tau,i)=1$.
We then merge the meaning classes as in \Eqn{eq:merge}.
Our studies are represented here by two simulations, S/T/C200 and S/L500-1/C200,
the latter including local borrowing. The per-capita death rate $\mu$ was set to a large value,
so that polymorhpism 
was low. 
We find, when we fit data of this kind, that the tree and its dates are robust to this form of model
mis-specification.

\subsection{Incorrect splitting deep rooted homology classes}
\label{sec:mmsplit}

When the scientist groups instances of traits into homology classes,
instances of traits born deep in the tree may be highly evolved,
and correspondingly difficult to identify as in fact homologous.
This error can populate the deeper branches
of the tree with spurious birth events. This is a case where model
mis-specification is not homogeneous over the tree, and will lead to over-estimation of the
tree depth. When we replace the \cite{ringe02} ``splitting'' data with the \cite{ringe02}
``lumping'' data we do see a 3\% downward shift in the estimated root time.

\subsection{Unknown vocabulary as absent traits}
\label{sec:mmzeros}
In our analysis of the \cite{ringe02} data, we retain some languages with
gaps, corresponding to missing data. We replace these gaps with zeros, marking trait absence.
Gappy languages (Hittite, Tocharian) do stand out in predictive tests counting singleton traits
on external data for the 328 and 200 word-lists. However, the effect is removed
when we reduce the data to the Swadesh 100 word-list, where traits are better attested.
The effect is to bias reconstructed branching times for gappy taxa to larger age values
on the \cite{ringe02} 328 and 200 word-list data (see for example {\sf HT} in \Fig{fig:cladeage}).
%

\section*{Acknowledgements}
The authors acknowledge advice and assistance from Quentin Atkinson and David Welch
of the University of Auckland, and financial support from the Royal Society of New Zealand.

\bibliographystyle{Chicago}
\bibliography{longlangpaper}
\end{document}